\documentclass[11pt]{amsart}
\pdfoutput=1
\usepackage{times}
\usepackage{comment}
\usepackage{epsfig}
\usepackage{varioref}
\usepackage{textcomp}
\usepackage{multirow}
\usepackage{graphics}
\usepackage{bmpsize}
\usepackage{url}
\usepackage{draftcopy}
\usepackage[numbers]{natbib}
\usepackage{amssymb}
\usepackage{algorithm}
\usepackage{algpseudocode}
\usepackage{varioref}
\usepackage{multirow}
\usepackage{enumitem}
\makeatletter\@ifclassloaded{beamer}{}{\usepackage{imakeidx}}\makeatother
\usepackage{lscape}
\usepackage{gensymb}
\usepackage[toc]{appendix}
\usepackage{mathtools}
\usepackage{pict2e}
\usepackage{picture}

\DeclareMathOperator*{\argmax}{arg~max}

\DeclareMathOperator*{\sign}{sgn}

\makeatletter\@ifclassloaded{beamer}{}{}\makeatother

\newcommand{\genlab}[2]{\label{#1:#2}}
\newcommand{\genref}[2]{#1~\vref{#1:#2}}

\newcommand{\eqnlab}[1]{\genlab{equation}{#1}}
\newcommand{\eqnref}[1]{\genref{equation}{#1}}

\newcommand{\figlab}[1]{\genlab{figure}{#1}}
\newcommand{\figref}[1]{\genref{figure}{#1}}

\makeatletter\@ifclassloaded{beamer}{}{%
    \newtheorem{theorem}{Theorem}

}
\makeatother
\newcommand{\thelab}[1]{\genlab{theorem}{#1}}
\newcommand{\theref}[1]{\genref{theorem}{#1}}

\newcommand{\pngfigure}[4][hbt]{\begin{center}\begin{figure}[#1]\includegraphics[angle=0,scale=#4]{#2.png}\caption[\hspace{\normalparindent}#3]{#3}\figlab{#2}\end{figure}\end{center}}
\makeatletter\@ifclassloaded{beamer}{\input{beamer}}{}\makeatother

\usepackage{booktabs}
\vrefwarning
\begin{document}
%
%
\title[Budget Threshold Utility]{Risk Limited Asset Allocation with a Budget Threshold Utility Function and Leptokurtotic Distributions of Returns}
\author{Graham L. Giller}
\email{graham@gillerinvestments.com}
\date{\today}
\begin{abstract}
An analytical solution to single-horizon asset allocation for an investor with a piecewise-linear utility function, called herein the ``budget threshold utility,'' and exogenous position limits is presented. The resulting functional form has a surprisingly simple structure and can be readily interpreted as representing the addition of a simple ``risk cost'' to otherwise frictionless trading.
\end{abstract}
\maketitle
\section{Introduction}
Beginning with the pioneering work of Markowitz\cite{harry1952markowitz}, the theory of optimal investment (and optimal trading) has been the subject of extensive analytical work, both in academia and within financial institutions. The purpose of this note is not to provide a comprehensive discussion of the subject, which can be found in numerous places such as the work of Ziemba and MacLean\cite{ziemba2017problems}, the recent book by Paleologo\cite{paleologo2025elements}, and the author's own works\cite{giller2023essays}. Nor is to provide a thoroughly developed theory of investment in theory or in practice, which can be found in many places including those cited, but merely to present an interesting\footnote{At least to the author.} analytical result that is of relevance to the general subject. Properly, this might be thought of as sitting somewhere between fully optimal investment strategy and a completely na\"ive investment strategy, as an informed reader will be able to readily identify structural flaws in some of the major propositions. In fact, a correspondent of the author went as far as to describe the shape of the utility function discussed as ``irrational.''\cite{viole2025private}. Nevertheless, the result presented is fully optimal \textit{given} the assumptions presented, connects a gap between risk neutral investment and risk averse investment, and is of a remarkably simple analytical form.
\section{The Budget-Threshold Utility Function}
Influenced by the work of Nawrocki and Viole\cite{viole2011utility}, and the author's personal experience in proprietary trading at a large investment bank\cite{giller2022adventures}, we construct a piecewise linear utility function that is indifferent to wealth, $W$\!, above a pre-defined threshold, or ``budget,'' $\beta$, and linear in the dis-utility of wealth below that budget. Mathematically, this can be written
\begin{equation}\eqnlab{utility}
U(W,\beta)=\begin{cases}
    \beta&W\ge\beta\\
    W&W<\beta
\end{cases},
\end{equation}
and is illustrated, for several values of $\beta$, in \figref{utility}, below.
\pngfigure{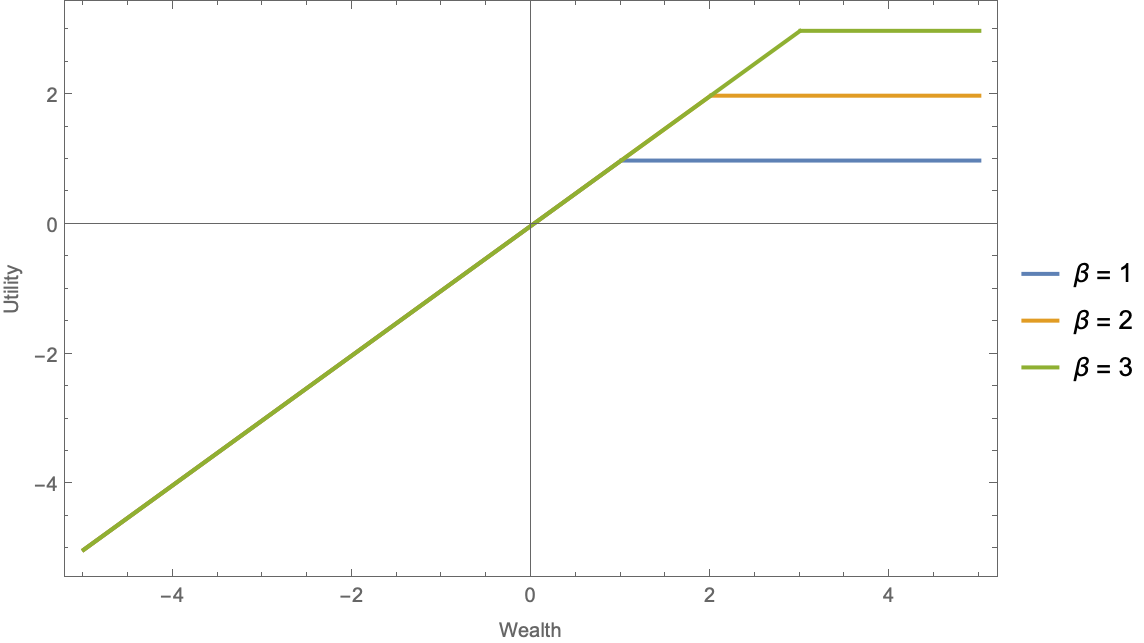}{Functional form of the budget-threshold utility function for several values of the budget, $\beta$.}{0.6}

Viole's criticism of this form\footnote{Viole, as previously cited} is that:
\begin{enumerate}[label=(\roman*)]
\item there is no ``credit'' received for wealth increases above the budget which, in an inter-temporal setting, could be used to decrease the instantaneous budget requirement for the next investment period while leaving the temporal average of the budget unchanged; and,
\item the linear dis-utility of losses does not sufficiently deter risk-taking behaviour when compared to, for example, quadratic functions such as those that appear in mean-variance optimization\cite{harry1952markowitz}.
\end{enumerate}
Nevertheless, this is a concave function that might be thought of as describing the objectives of ``some'' investors or traders, and is interesting to investigate.
\section{The Distribution of Returns}
The distribution of asset returns over the investment horizon considered is taken to be represented by the \textit{Generalized Error} distribution\footnote{Sometimes called the Generalized Normal distribution or the Exponential Power distribution.}, which is a symmetric univariate distribution from the exponential family\cite{forbes2011statistical} that may be parametrically deformed into a range of shapes including both the Laplace distribution and the Normal distribution, as well as more leptokurtotic varieties and platykurtotic varieties up to a limiting form as the Uniform density. This choice is motivated by the author's personal research into describing the distribution of returns in securities markets that are consistent over extensive periods of history\cite{giller2022adventures}.

The $\mathrm{GED}$ has many parameterizations, but the one used here is:
\begin{equation}\eqnlab{ged}
    r\sim\mathrm{GED}(\mu,\sigma,\kappa)\Leftrightarrow f(r|\mu,\sigma,\kappa)=\frac{
    e^{-\frac{1}{2}\left|\frac{r-\mu}{\sigma}\right|^{\frac{1}{\kappa}}}
    }{2^{\kappa+1}\sigma\Gamma(\kappa+1)}
\end{equation}
where $f(\cdot)$ is the probability density function of $r$ given parameters $(\mu,\sigma,\kappa)$. These specify the location, scale, and kurtosis\footnote{The kurtosis is a \textit{function} of $\kappa$ and the standard deviation is a function of $\sigma$ and $\kappa$, see Giller\cite{giller2005generalized}.} of the distribution of returns, respectively. This form is chosen \textit{specifically} so that the Normal limit ($\kappa=1/2$) appears naturally, in its common form, without a required reparameterization of $\sigma$. 
If $r$ is taken to be a major market index, such as the S\&P~500, it is found that, for most of recorded history:\footnote{i.e. From 1929--2025.} $\mu>0$; $\sigma$ follows some form of asymmetric GARCH process\cite{glosten1993relation,engle2001garch}; and $\kappa$ is around $0.75$\cite{giller2022adventures}.
\section{Choice of Budget and Expected Utility}
To find an optimal holding, $\hat{h}$, in an asset with some expected return, $\alpha$, it is necessary to compute the expected utility under the given distribution of returns and then maximize the resultant function. i.e.
\begin{equation}
    \hat{h}=\argmax_h\mathbb{E}[U(W(h),\beta)|\alpha].
\end{equation}
To obtain a solution, the following choices are made:\footnote{Equivalently, the entire discussion can be framed in terms of \textit{marginal} utility without any required changes.}
\begin{enumerate}[label=(\roman*)]
    \item the initial wealth is zero, thus the wealth in the utility expression is solely a function of the incremental profits due to investment;
    \item the future wealth, given $h$ and $r$, is then simply their product, $hr$; and,
    \item the budget is taken to be the expected value of the future wealth, or $h\alpha$.
\end{enumerate}
Then
\begin{equation}\eqnlab{expute}
    \mathbb{E}[U](h|\alpha,\sigma,\kappa)=\int_{-\infty}^\infty U(hr,h\alpha)f(r|\alpha,\sigma,\kappa)\,dr.
\end{equation}

As the utility is piecewise linear, it is immediately obvious that, in the framework of Nawrocki and Viole, this function is the sum of the lower-partial moment\footnote{The conventional statistical moment of order $n$ about $a$, $\mu_n'(a)=\mathbb{E}[(x-a)^n]$, may be partitioned into the sum of a lower-partial moment, $l_n'(a)=\mathbb{E}[(x-a)^n|x\le a]$, and an upper-partial moment, $u_n'(a)=\mathbb{E}[(x-a)^n|x>a]$.} of order $1$ and the upper-partial model of order $0$, both evaluated with reference to the expected profit, $h\alpha$. i.e.
\begin{equation}    
\mathbb{E}[U]=l_1'(h\alpha)+u_0'(h\alpha)
\end{equation}
Furthermore, due to the linearity, this integral exists and may be evaluated. It may be shown to be
\begin{align}
    \mathbb{E}[U](h|\alpha,\sigma,\kappa)&=h\left(\alpha-\sigma\tau(\kappa)\sign h\right)\eqnlab{expute2}\\
    \mathrm{where}\;\tau(\kappa)&=\frac{\Gamma(2\kappa+1)}{2^{2-\kappa}\Gamma(\kappa+1)}\eqnlab{taufun}.
\end{align}

Although, as a ratio of Gamma functions, $\tau(\kappa)$ is strongly divergent, within the region of interest, $1/2\le\kappa\le1$, it is remarkably well behaved, growing relatively slowly from $1/\sqrt{2\pi}$ at the lower edge of this range to just $1$ at the upper edge.
\pngfigure{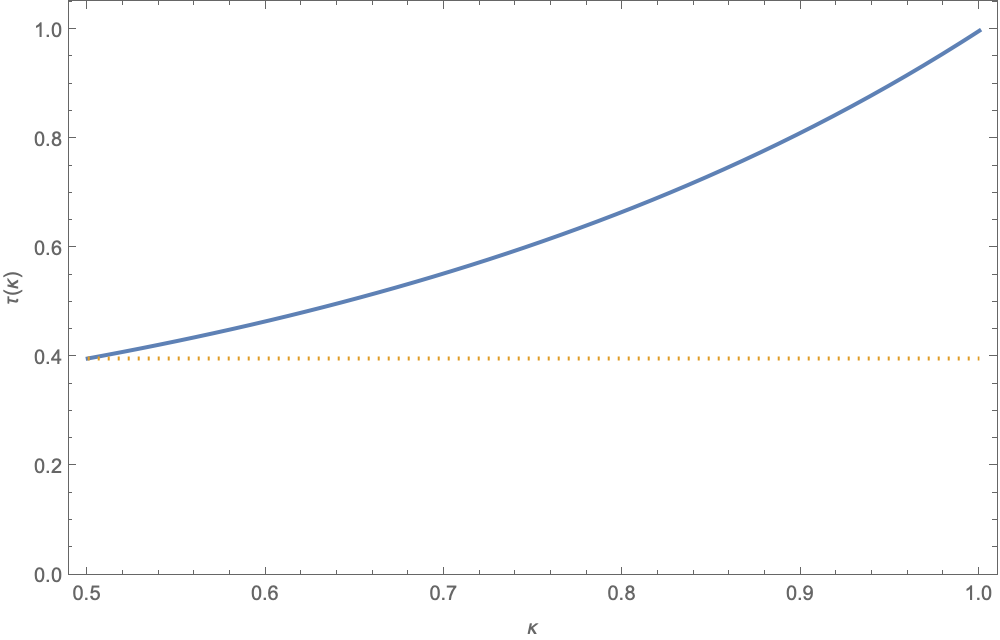}{The risk scaling function over the region of interest.}{0.6}
\section{Maximum Expected Utility Subject to an Exogenous Risk-Limit}
With asymptotically linear utility functions an obvious defect exists: there is no upper or lower limit to the positions taken that arises \textit{naturally} from the problem specification. This is unlike mean-variance optimization in which the quadratic form guarantees a finite solution for finite arguments. However, this defect can be regularized away by introducing an exogenous risk limit, $L>0$, to the problem. i.e.
\begin{equation}
    \hat{h}=\argmax_h\mathbb{E}[U](h|\alpha,\sigma,\kappa)\;\mathrm{becomes}\;\hat{h}=\argmax_{h\in[-L,+L]}\mathbb{E}[U](h|\alpha,\sigma,\kappa).
\end{equation}
There is no practical loss of generality to this modification of the problem as \textit{all real investors face such constraints}, whether they arise naturally from their finite wealth and credit prospects or from an institutional risk-management role. With this regularization step, the optimal position is readily obtained and is given in \theref{solution}.
\begin{theorem}\thelab{solution}
The position for a risk-limited investor with a budget threshold utility function that maximizes the expected utility when returns are drawn from a Generalized Error distribution is given by:
\begin{equation}\eqnlab{solution}
    \hat{h}=\begin{cases}
    L\sign\alpha&|\alpha|>\sigma\tau(\kappa)\\
    0&\mathrm{otherwise}
    \end{cases}.
\end{equation}
\begin{proof}
Consider $h>0$: in this circumstance \eqnref{expute2} becomes, simply,
\begin{equation}
    \Omega(h)=\mathbb{E}[U](h|\alpha,\sigma,\kappa)=h\left(\alpha-\sigma\tau(\kappa)\right)\;\Rightarrow\;\left.\frac{d\Omega}{dh}\right|_{h>0}=\alpha-\sigma\tau(\kappa).
\end{equation}
It follows that for any given $h>0$ such that $\alpha>\sigma\tau(\kappa)$ it is always true that there exists $h'\in(h,L]:\Omega(h')>\Omega(h)$. Thus, if $\alpha>\sigma\tau(\kappa)$ then $\hat{h}=+L$.
Now consider $h>0$ but $0\le\alpha\le\sigma\tau(\kappa)$: in this circumstance $\Omega(h)\le0$ but we observe that $\Omega(0)=0$. Thus it is sub-optimal to chose a non-zero position when $0<\alpha\le\sigma\tau(\kappa)$, or $\hat{h}=0$.

Consider $h<0$: complementary arguments clearly lead to the same conclusions for negative alphas but with the sign exchanged.
Finally, consider $\alpha=0$: for this case $\Omega(h)<0\;\forall\; h\ne0$, therefore $\hat{h}=0$.
\end{proof}
\end{theorem}
\section{Interpretation of the Result and\\ a Practical Algorithm for Traders}
I believe that, despite the apparent na\"ivety of the problem proposed, the solution is, in fact, quite interesting. It is well known that a risk-limited gross-profit maximizing investor should obtain a position as large as possible in the direction of the alpha\cite{giller2023essays}. The introduction of the budget threshold, as specified in this work, changes such a long-short, or ``binary,'' trading algorithm to a three state system that considers long, short, and flat positions only. The term $\sigma\tau(\kappa)$ can be thought of as a ``risk cost'' that must be exceeded to make risk-taking worthwhile to the investor. 

This risk cost is a function of the parameterization of the distribution of returns, albeit not a particularly strongly varying function. For returns described by \eqnref{ged}, the variance of returns is given by
\begin{equation}\eqnlab{var}
s^2=\mathbb{V}[r]=\frac{2^{2\kappa}\Gamma(3\kappa)}{\Gamma(\kappa)}\times\sigma^2.
\end{equation}
Expressing the risk cost in terms of $s$, the standard deviation of returns, gives
\begin{equation}
    \sigma\tau(\kappa)=\frac{1}{4}\sqrt{\frac{\Gamma(\kappa)}{\Gamma(3\kappa)}}\frac{\Gamma(2\kappa+1)}{\Gamma(\kappa+1)}\times s.
\end{equation}
It can clearly be seen from \figref{risk} that the leptokurtosis of the distribution of returns \textit{barely} affects the scaling of the standard deviation, $s$, in the formula for the size of the risk cost ``barrier.'' Thus this is a clear example of a circumstance in which assuming Normally distributed returns \textit{doesn't} actually damage the optimal policy framework.
\pngfigure{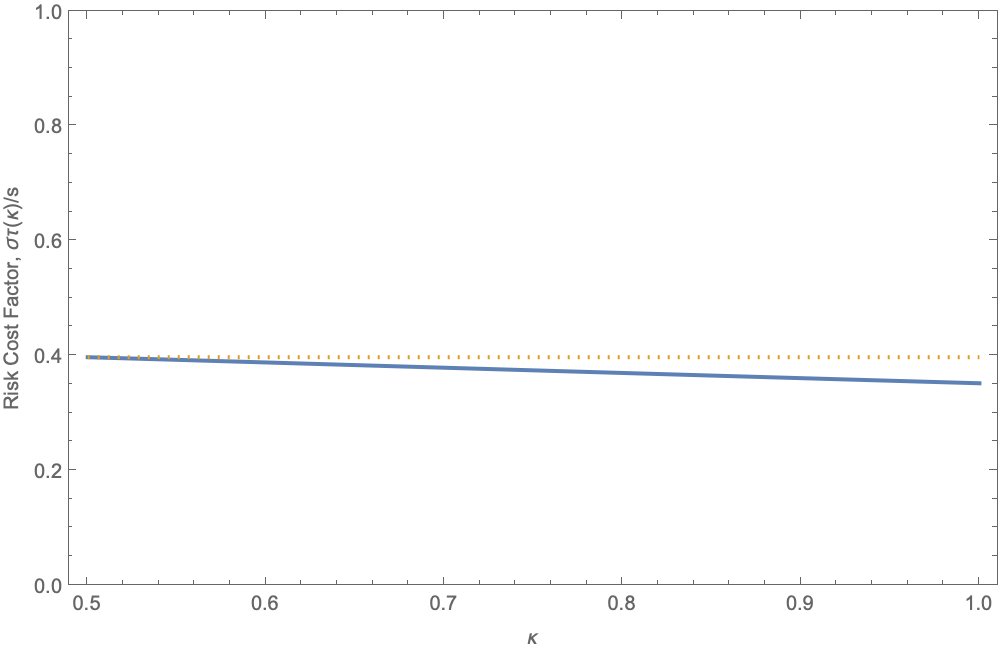}{Scaling of the risk cost factor with the kurtosis parameter of the distribution of returns over the region of interest. The ``Normal distribution theory'' value is represented by the dotted line.}{0.6}

Nevertheless, given the commonly acknowledged non-stationarity of financial distributions and the simplicity of the optimal algorithm, it seems well motivated to follow a ``semi-empirical'' approach, similar to that adopted in Nuclear Physics\cite{bethe1936semf,weizsacker1935semf}, and replace this parametric expression with a simple constant, $K$, to be determined from empirical optimization (i.e. backtesting). A practical algorithm for traders is then
\begin{equation}\eqnlab{sesolution}
    \hat{h}=\begin{cases}
    L\sign\alpha&|\alpha|>Ks\\
    0&\mathrm{otherwise}
    \end{cases},
\end{equation}
where $\alpha$ and $s$ are the mean and standard deviation of future returns under the conditional distribution used by a trader for a given asset, and $K\approx0.4$. This trading algorithm, in the form of a ``holding function,''\footnote{Literally a function that tells the trader want position to hold in response to their inputs.} is illustrated in \figref{algo}.
\pngfigure{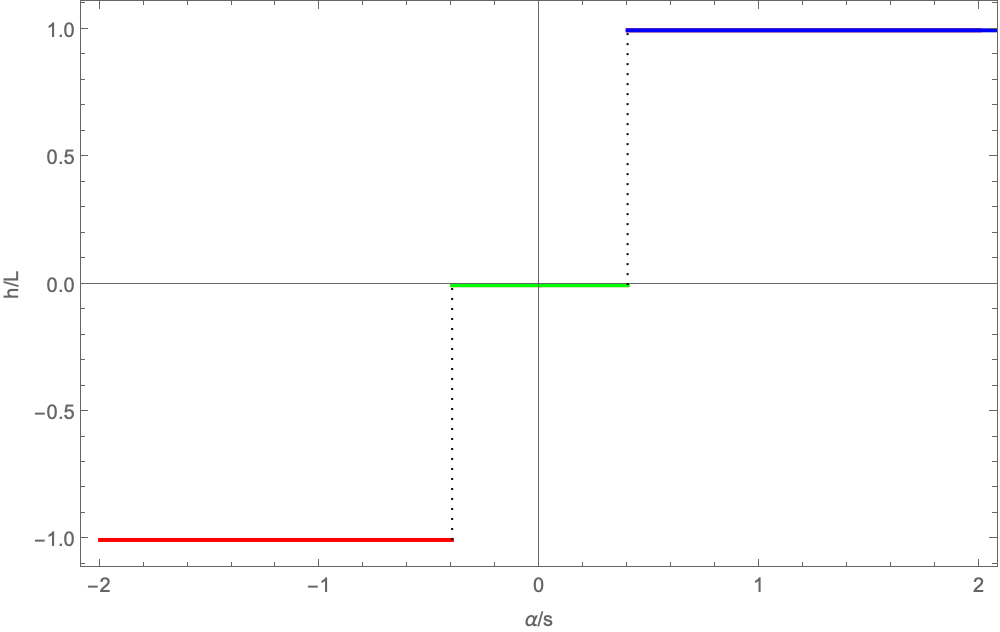}{The ``holding function'' corresponding to the optimal strategy for a trader with a budget threshold utility function expressed as the relative position size as a function of the standardized alpha.}{0.6}
%
%
\bibliographystyle{plain}
\bibliography{citation}
\end{document}